\documentclass[journal]{IEEEtran}
\ifCLASSINFOpdf
\else
\fi

\usepackage{enumerate}
\usepackage{mdwlist}
\usepackage{eurosym}
\usepackage{graphicx}
\usepackage{color}
\usepackage{amssymb,amsmath}
\usepackage{algorithm2e}
\usepackage{ctable}
\usepackage{multirow}
\usepackage{rotating}
\usepackage{flushend}
\usepackage{cite}
\usepackage{booktabs}
\usepackage{url}
\usepackage[caption=false,font=footnotesize]{subfig}

\usepackage{tablefootnote}
\usepackage[bottom]{footmisc}

\begin{document}
%
\title{Co-optimization of Energy and Reserve with Incentives to Wind Generation: Case Study}
%
%
%

\author{Yves~Smeers, 
    Sebastian~Martin,~\IEEEmembership{Member,~IEEE,}        
	and Jos\'e A.~Aguado,~\IEEEmembership{Member,~IEEE}
\thanks{Y. Smeers is with the Center for Operations Research and Econometrics (CORE) at Universit\'e Catholique de Louvain
(UCL), Belgium (e-mail: yves.smeers@uclouvain.be)}
\thanks{S. Martin and J. A. Aguado are with the Department of Electrical Engineering, University of Malaga, Malaga, Spain (e-mail:
smartin@uma.es; jaguado@uma.es). Their work was supported in part by the Spanish Ministry of Economy and Competitiveness through
the project ENE2016-80638-R, and in part by the University of M\'alaga (Campus de Excelencia Internacional Andaluc\'ia Tech).}%
}

\maketitle

\begin{abstract}
This case study presents an analysis and quantification of the impact of the lack of co-optimization of energy and reserve in the presence of high penetration of wind energy. The methodology is developed in a companion paper, Part I. Two models, with and without co-optimization are confronted. The modeling of reserve and the incentive to renewable as well as the calibration of the model are inspired by the Spanish market. A sensitivity analysis is performed on configurations that differ by generation capacity, ramping capability, and market parameters (available wind, Feed in Premium to wind, generators risk aversion, and reserve requirement). The models and the case study are purely illustrative but the methodology is general. 
\end{abstract}


%
\IEEEpeerreviewmaketitle

\setlength{\abovecaptionskip}{0pt plus 0pt minus 2pt}
\setlength{\belowcaptionskip}{0pt plus 0pt minus 2pt}
\setlength{\textfloatsep}{0pt}
\setlength{\intextsep}{0pt}
\section*{Notation}
Only the terms used in this Part II are included in this section. For a complete list of terms see Part I.

\subsection{Indices and sets}
\begin{basedescript}{\desclabelstyle{\pushlabel}\desclabelwidth{1cm}}
 \item[$g, G$] Dispatchable generators, $g \in G$.
 \item[$w, WT$] Wind turbines, $w \in WT$.
 \item[$k, \Omega$] Index and set for scenarios, $k \in \Omega$.
\end{basedescript}

\subsection{Parameters}
\begin{basedescript}{\desclabelstyle{\pushlabel}\desclabelwidth{0.8cm}}
 \item[$\overline{A} / \underline{A}$] Upper/lower factor for required up. reserve (p.u.).
 \item[$B$]Lower bound factor for committed downward reserve respect to the committed upward reserve (p.u.).
 \item[$\overline{R}_g / \underline{R}_g $] Upward/Downward ramping slope respect to the generation capa\-city (p.u.).
 \item[$C_g$] Slope of gen. cost for dispatch. gen. $g$, (\officialeuro/MWh).
 \item[$FiP$] Premium to the schedu. wind gen., (\officialeuro/MWh).
 \item[$M_y$] Balancing reserve factor for wind turbines, (p.u).
 \item[$M_x$] Balancing reserve factor for dispatch. gen., (p.u.).
 \item[$\Gamma_0$] Constant of the inverse demand function, (\officialeuro/MWh).
  \item[$\Theta$] Confidence level for firm Conditional Value at Risk (CVaR), $\Theta \in (0,1)$.
 \item[$\Lambda$] Parameter for ramping availability, $\Lambda \in [0,1]$.
 \item[$\Phi_0$] Slope of the inverse demand function (\officialeuro/(MWh)$^2$).
 \item[$\Psi$] Level of risk aversion, $\Psi \in [0,1]$, zero is risk neutral. 
\end{basedescript}

\subsection{Variables}
\textbf{Primal variables}
\begin{basedescript}{\desclabelstyle{\pushlabel}\desclabelwidth{1.22cm}}
 \item[$d$] Energy sales of the firm in day-ahead, (MWh).
 \item[$ru_g / rd_g$] Committed upward/downward reserve capacity from dispatchable generator $g$, (MW).
 \item[$x_g$] Scheduled generation of dispatchable unit $g$, (MW).
 \item[$y_w$]  Scheduled generation of wind turbine $l$, (MW).
\end{basedescript}

\textbf{Dual variables (all in \officialeuro/MWh)}
\begin{basedescript}{\desclabelstyle{\pushlabel}\desclabelwidth{1cm}}
 \item[$\overline{\gamma} / \underline{\gamma}$] Upper bound committed upward/downward reserve.
 \item[$\overline{\kappa} / \underline{\kappa}$] Lower bound for committed up./down. reserve.
\end{basedescript}


\section{Case Study: Description and Data}\label{s:Case_Study_Description}

The case study is  inspired by a phenomenon that affected the dynamics of European generation in the period 2008-2013 \cite{500_Billions_The_Economist} and still persists in moderate form today \cite{IEEFA_2017, 2020_EC_Energy_Price, 2020_EC_Q4_Energy_Price}. Existing flexible Combined Cycle Gas Turbines (CCGT) necessary to the system because of their contribution to reliability and flexibility became unprofitable, with the consequence that special non market arrangements had to be found to retain them in the system. This situation was summarized in \cite{2013_Agora}: {\it ``There's plenty of flexibility, but so far it has no value''}. The statement can be interpreted in two ways: either the system is awash with flexibility and its short-term price is effectively null, which is a transient phenomenon; alternatively the market does not price flexibility in a correct way, which is a matter of market design. We use the models COM (Co-Optimization) and EQM (Equilibrium Model) developed in Part I to assess whether co-optimization of energy and reserve would have significantly affected the cash flow generated by these flexible units. The analysis is illustrative but the methodology can easily be scaled up and is only limited by the capabilities of commercial optimization codes to solve the underpinning two stages stochastic programs.

We structure the analysis around two questions:
\begin{enumerate}
 \item Can one characterize a range of situations where flexibility, expressed in terms of reserves requirements, has effectively no value and alternatively, conditions where it has value? If so, can one characterize situations that retain flexible equipment in the system? 
 \item Does market design (co-optimization) have an impact on the above? And if so, how does it depend on market parameters such as wind condition $\mu$, Feed in Premium $FiP$ to wind, generators' risk aversion $\Psi$, and reserve requirement $M_y$?
\end{enumerate}

This section \ref{s:Case_Study_Description} of Part II introduces  a case study that considers a number of configurations of generation capacity (High, Low, and Very Low), ramping capacity (High, Low), and market parameters (wind availability $\mu$, $FiP$, risk aversion $\Psi$, reserve requirement $M_y$). Section \ref{s:Case_Study_Results} contains a discussion of the numerical results. The conclusions in Section \ref{s:Conclusion} close the paper.

\subsection{Energy demand}\label{ss:energydemand}

The price and energy demand are endogenous and related by a linear inverse demand function, \eqref{eq:InverDeman}:
\begin{flalign}
 & \textrm{Price} = \Gamma_0 - \Phi_0 \!\cdot\! (\textrm{Energy Demand in Day-Ahead}) & \label{eq:InverDeman}
\end{flalign}
where $\Gamma_0 = 209.78$ (\officialeuro/MWh) and $\Phi_0 = 0.0056$ (\officialeuro/MW$^2 \cdot $h), based on a calibration on historical data from the Spanish system. The confidence level for Conditional Value at Risk (CVaR) is $\Theta = 0.95$.

\subsection{Generation, Ramping and Market Configurations}\label{ss:GenRamp}

\setlength\tabcolsep{4pt}
\begin{table}[t]
\caption{Summary of generation configurations}
\begin{center}\label{tab:GeneratorCaseStudy}
\footnotesize
\begin{tabular}{c|p{0.8cm}rrrrrr}
 &  & High Gen. &  Low Gen. & Very Low & $C_g$  &  $\overline{R}_g \!=\! \underline{R}_g$ \\
&   &  $\overline{X}_g$  &  $\overline{X}_g$ &   $\overline{X}_g$  &   &   \\
$g$ & Tech.  &  (MW)  &   (MW) &   (MW)  &  (\officialeuro/MWh)  &  \% of $\overline{X}_g$ \\
\hline 1 & CCGT & 0 & 3274.98 & 1274.98 & 44.31 & 53.33 & \\
2 & CCGT & 0 & 2056.58 & 2056.58 & 43.88 & 53.33 & \\
3 & CCGT & 10632.7 & 2153.5 & 2153.5 & 43.45 & 53.33 & \\
4 & Nuclear & 1519.23 & 1519.23 & 1519.23 & 10.91 & 2.08 & \\
5 & Nuclear & 6053.35 & 6053.35 & 6053.35 & 10.29 & 2.08 & \\
6 & Coal & 2035.89 & 2035.89 & 2035.89 & 37.5 & 20 & \\
7 & Coal & 5119.13 & 5119.13 & 5119.13 & 38.44 & 25 & \\
8 & Coal & 1198.12 & 1198.12 & 1198.12 & 19.77 & 25 & \\
9 & Coal & 1945.51 & 1945.51 & 1945.51 & 20.24 & 25 & \\
\hline &  Wind  & 22573.00 &  22573.00 & 22573.00 & - & - \\
 & Dispatch. & \textbf{28503.93} &  \textbf{25356.29} & \textbf{23356.29} & - & - \\
\hline & \textbf{Total}  & 51076.93 &  47929.29 & 45929.29 & - & -
\end{tabular}
\end{center}
\end{table}

Three configurations of generation capacity are considered and presented in Table \ref{tab:GeneratorCaseStudy}. Wind capacity (22.57 GW) and the demand function are identical in all cases. Dispatchable generation capacities take three values: 28.5 GW (High), 25.35 GW (Low), and 23.35 GW (Very Low). These values can be interpreted as stages in the retirement of CCGT to restore profitability of the remaining capacity.

The parameter $\Lambda \in [0,1]$ is described in detail in Part I. It is used for testing the capability of the system to manage forecast errors on wind generation. $\Lambda$ sets the ramping capacity already committed in the scheduled generation to move from a period to the next one. For instance, for the upward reserve the capacity already committed (not available for flexibility) is $\overline{R}_g \!\cdot\! \Lambda \!\cdot\! x_g$. Two configurations for ramping capability are considered: Low $\Lambda = 1$, and High $\Lambda = 0$. $\Lambda$ takes the same value for all dispatchable generators in each test. 


The characteristics of the six possible combinations of generation and ramping and the number of the figures reporting the results are given in Table \ref{tab:GenAndRamping}. The reference situation is taken as Low ramping with Low generation. We also consider configurations with low generation and high ramping to develop a story line. 

Four parameters further characterize the market configurations: wind condition $\mu$, $FiP$ to wind, generators' risk aversion $\Psi \in [0,1]$, and reserve requirement $M_y$ by renewable generation. Eight configurations, listed on Table \ref{tab:ParametersValues}, are considered in the case study. We refer to those configurations in the figures by using the abbreviated name in the columns' head in Table \ref{tab:ParametersValues}.


\setlength\tabcolsep{2pt}
\begin{table}[b]
\caption{Summary of data for generation and ramping configurations}
\begin{center}\label{tab:GenAndRamping}
\footnotesize
\begin{tabular}{l|c|ccc}
 Config. & Fig. & Disp. Gen. & Wind Gen. & Ramp. ($\Lambda$) \\
 &  & (GW) & (GW) & (p.u.) \\
\hline HH (High Gen. High Ramp.) & \ref{fig:HH} & 28.50 & 22.57 & 0.00  \\
HL (High Gen. Low Ramp.) & \ref{fig:HL} & 28.50 & 22.57 & 1.00\\
LL (Low Gen. Low Ramp.) & \ref{fig:LL} & 25.35 & 22.57 & 1.00\\
VLL (Very Low Gen. Low Ramp.) & \ref{fig:VLL} & 23.35 & 22.57 & 1.00\\
VLH (Very Low Gen. High Ramp.) & \ref{fig:VLH} & 23.35 & 22.57 & 0.00
\end{tabular}
\end{center}
\end{table}

\begin{table}[b]
\caption{Summary of values for market configurations}
\begin{center}\label{tab:ParametersValues}
\begin{tabular}{l|rrrrrrrr}
& Wind & Wind & $FiP$ & $FiP$ & Risk & Risk & Res. & Res. \\
&  L. & H. & L. & H. & L. & H. & L. & H. \\
\hline $\mu$, wind (\%)  & 4.01 & 65.00 & 23.83 & 23.83 & 23.83 & 23.83 & 23.83 & 23.83 \\
$FiP$, (\officialeuro/MWh) & 30.00 & 30.00 & 0.00 & 80.00 & 30.00 & 30.00 & 0.00 & 80.00 \\
$\Psi$, risk avers. & 0.40 & 0.40 & 0.40 & 0.40 & 0.00 & 1.00 & 0.40 & 0.40 \\
$M_y$, (\%) & 15.00 & 15.00 & 15.00 & 15.00 & 15.00 & 15.00 & 60.00 & 60.00
\end{tabular}
\end{center}
\end{table}

\subsection{Reserve Modeling}\label{ss:ReserveModeling}

Reserve is modeled using what we call balancing reserve factors (MW of reserve/MW scheduled), \cite{HEGGARTY20191327}, $M_x = 0.02$ for conventional generators in all the configurations, and $M_y$ for wind generators. The values of these parameters are calibrated on the historical observations in the Spanish market. For wind turbines we consider two cases that reflect the error forecast due to the time between commitment and real time:
\begin{itemize}
 \item $M_y = 0.15$, for short time horizon, around 6 h.
 \item $M_y = 0.60$, for long time horizon, around 24 h.
\end{itemize}
Longer forecasting time horizon imply greater uncertainty and forecast errors.

Using the balancing factors, the reserve required by the Transmission System Operator (TSO) is given by \eqref{eq:ReservReq},
\begin{flalign}
 & M_x \!{\cdot}\! \sum_{g \in G} x_g + M_y {\cdot} \sum_{w \in WT} y_w & \label{eq:ReservReq}
\end{flalign}
where $x_g$ is the scheduled dispatchable generation by $g$ and $y_w$ is the scheduled wind generation by turbine $l$. The constraints for the upward, $\sum_{g \in G} ru_g$, and downward, $\sum_{g \in G} rd_g$, reserves are also inspired by the Spanish Grid Code, \cite{2006_Reserva_frecuencia_Potencia}. The upward reserve must remain between 90\%, $\underline{A} = 0.9$, and 110\%, $\overline{A} = 1.1$, of the reserve required by the TSO, \eqref{eq:ReservReq}. The downward reserve must remain between 40\%, $B = 0.4$, and 100\% of the value of the upward committed reserve. The value for the committed reserves corresponds here to the sum of the primary and secondary reserves.


\subsection{Uncertain Parameters}
The available wind generation is the only uncertain parameter in the analysis. The values of $\mu$ and the scenarios are reported in Table \ref{tab:ScenarioTree}; all these values are given as a percentage of the rated wind generation capacity, that is 22573 MW, as indicated in Table \ref{tab:GeneratorCaseStudy}.  We consider the wind variability through the expected value of wind $\mu$, and also the forecast errors through 12 scenarios for each value of $\mu$. We consider three values of $\mu$ that correspond to low wind $\mu = 4.01\%$, average wind $\mu = 23.83\%$, and high wind $\mu = 65.00\%$. All scenarios have the same probability, $\frac{1}{12} \approx 8.33$\%.  The following describes the construction of the wind scenarios taking into account the forecast errors.

We use a Beta distribution to model the wind power forecast error. Let the load factor for wind generation be $q = \frac{\textrm{Power output}}{\textrm{Rated power}} \in [0,1]$ , then according to \cite{2002_Bofinger, 2005_Fabbri} this load factor fits a Beta distribution: $f(q) = \frac{\Gamma(\alpha + \beta)}{\Gamma(\alpha)\Gamma(\beta)}q^{\alpha-1}\!\cdot\!(1 - q)^{\beta-1}$, $q \in [0,1]$, 
 where $\frac{\Gamma(\alpha + \beta)}{\Gamma(\alpha)\Gamma(\beta)}$ is a scale factor such as $\int_0^1 f(x)dx = 1$, and the parameters $\alpha$ and $\beta$ are directly related with the mean $\mu$ and the standard deviation $\sigma$ of that distribution: $\alpha = \mu^2 \frac{1 - \mu}{\sigma^2} - \mu$ and $\beta = \alpha \!\cdot\!\left(\frac{1}{\mu} - 1 \right)$.
 
The analysis of empirical data shows that $\sigma$ fits a linear function of $\mu$, $\sigma(\mu) = k_1 \!\cdot\!
\mu + k_2$, \cite{2002_Bofinger, 2005_Fabbri, 2009_Ortega}, where the coefficients $k_1$ and $k_2$ depend mainly on
the time horizon and the geographic dispersion of the wind turbines. Here we use the expression given in \cite{2009_Ortega} for large scale generation and a time horizon of 24 h, and assume the same expression for a horizon of 6h: $\sigma = \frac{1}{5}\mu + \frac{1}{50}$ (in per unit).

To build the scenarios, we divide the range $[0,1]$ for the load factor into segments and associate each scenario with a segment. Here we consider 12 scenarios, let $k$ be the index for the extreme points of each segment, $z_k \in [0,1]$, then the range $[0,1]$ is discretized using
13 points, $0 = z_1 < z_2 < \ldots <z_{13} = 1$. The value and the probability of scen. $k$ are:
\begin{enumerate}[a)]
 \item Value of scenario $k$: $\mu(k) = \int_{z_k}^{z_{k+1}} \frac{\Gamma(\alpha +
\beta)}{\Gamma(\alpha)\Gamma(\beta)}x^{\alpha}\!\cdot\!(1 - x)^{\beta-1} dx$, that is the expected value on the
segment that defines the scenario.
 \item Probability of scenario $k$, $pr(k) = \int_{z_k}^{z_{k+1}} \frac{\Gamma(\alpha +
\beta)}{\Gamma(\alpha)\Gamma(\beta)}x^{\alpha-1}\!\cdot\!(1 - x)^{\beta-1} dx$, integral of the
probability density function of the Beta distribution on the segment associated with the scenario.
\end{enumerate}

The points $0 = z_1 < z_2 < \ldots < z_{13} = 1$ are selected to get segments of equal probability: $\frac{1}{12} = \int_{z_k}^{z_{k+1}} \frac{\Gamma(\alpha + \beta)}{\Gamma(\alpha)\Gamma(\beta)}x^{\alpha-1}\!\cdot\!(1 - x)^{\beta-1} dx$,
$k=1,2,\ldots,12$.

\setlength\tabcolsep{6pt}
\begin{table}[t]
\caption{Scenarios for available wind (\% on rated capacity)}
\begin{center}\label{tab:ScenarioTree}
\small
\begin{tabular}{p{0.5cm}|rrr||p{0.5cm}|rrr}
      Scen. & \multicolumn{3}{c||}{ $\mu$ (\%), (base 22.57 GW)} & Scen. & \multicolumn{3}{c}{ $\mu$ (\%), (base 22.57 GW)} \\
\cline{2-4} \cline{6-8} $k$ & 4.01 & 23.83 & 65.00 & $k$ & 4.01 & 23.83 & 65.00 \\
\hline 1 & 0.59 & 12.64 & 35.55 & 7 & 3.66 & 24.09 & 67.81 \\
2 & 1.20 & 16.13 & 46.56 & 8 & 4.27 & 25.60 & 71.13 \\
3 & 1.69 & 18.12 & 52.44 & 9 & 5.00 & 27.27 & 74.52 \\
4 & 2.15 & 19.76 & 56.99 & 10 & 5.93 & 29.24 & 78.18 \\
5 & 2.62 & 21.24 & 60.89 & 11 & 7.27 & 31.86 & 82.46 \\
6 & 3.12 & 22.66 & 64.44 & 12 & 10.57 & 37.34 & 89.05
\end{tabular}
\end{center}
\end{table}

\subsection{Models Implementation and Solving}

COM is a quadratic programming problem with linear constraints, made of 367 equations and  284 variables. As expected it is solved with no difficulty by standard off the shelf solvers such as CPLEX, CONOPT and MINOS under GAMS \cite{GAMS_24_1_3}. The computation time is around 0.78 seconds per problem on a machine with Intel i7-5820K CPU @ 3.30 GHz and 32 GB RAM under Debian 4.19.171-2 x86\_64. We crosschecked the model with two alternative formulations using the Karush-Kuhn-Tucker ( KKT) conditions of COM, and solving it with PATH, also under GAMS. One formulation with  handwritten KKT, and the other provided by the automatic generation of the dual by the Extended Mathematical Programming on GAMS. We got the same results with the three formulations.

EQM is a linear complementarity problem that modifies COM's KKT conditions (see Part I). This problem misses the perfect arbitraging between reserve and energy of COM. An interesting question is whether this is reflected in solving capabilities. And indeed a direct application of PATH (under GAMS) on EQM failed in some cases while PATH never failed on the corresponding COM problems. We accordingly used an iterative approach that interacts PATH with a linear sub-problem calculating the CVaR to update the risk neutral probabilities at each iteration. This worked in all studied cases, but it is at this stage a heuristic. Each EQM problem is made of 636 equations and 636 variables, and each linear subproblem for the CVaR contains 14 equations and 13 variables. Computation time to solve each EQM using the iterative approach is $\approx$ 1.86 sec. using GAMS \cite{GAMS_24_1_3} on a machine with Intel i7-5820K CPU @ 3.30 GHz and 32 GB RAM under Debian.

COM is a convex problem that, except for degeneracy, always has an unique solution. Scalability is not an issue as the (very high) capabilities of LP commercial codes set the limits. The situation is different for EQM that is not amenable to optimization. One needs to verify the existence of an equilibrium and the possibility of multiple solutions; one must also examine scalability. A full answer to these questions goes beyond the scope of this paper but the following gives the intuition.

Existence of a solution to EQM can be proved by a homotopy argument, \cite{2010_Hering}. One can show that a solution of COM is a solution of a modified EQM, and that the EQM that one wants to solve can be obtained by a continuous transformation of COM. A standard reasoning of degree theory will then imply that the uniqueness of the solution of COM implies a finite number of isolated solutions of EQM. These mathematical properties have an economic interpretation. As argued in Part I, COM differs from EQM by the fact that co-optimization takes into account the opportunity cost on reserves of decisions of energy. The continuous transformation of COM into EQM amounts to progressively decrease the impact of this opportunity cost. Multiplicity of solution also has an economic interpretation: co-optimization tries to minimize the excess demand (positive or negative) by a search of both energy and reserve prices. It is easy to show that this excess demand is a ``monotone'' (that is well behaved) mapping of these energy and reserve prices. We argued in Part I that the separation of energy and reserve (dropping co-optimization) implies that one needs to find a zero of the excess demand by only playing with reserve prices, and letting the market find the corresponding energy production and price through a separate auction. The problem is that the obtained excess demand of reserve is not necessarily monotone (can be badly behaved) in the price of reserve, which is the source of a multiplicity of solutions. This has consequences on scalability. There is now a lot of experience (that originated in Hogan's work on project independence in the seventies \cite{HOGAN1975251}) in solving these problems by sequences of optimizations problems. Except for the need to resort to an iterative procedure, scalability is only restricted by the possibilities offered by commercial codes. But this remains a heuristic.

A final question is whether multiplicity of solution can occur in practice. As argued later in the case study multiplicity of solutions can indeed occur. Former European Union (EU) experience with the separation of the clearing energy and transmission showed that awkward situations could happen in practice where the clearing of transmission and energy were incompatible (transmission rights and energy flows were in opposite directions). This pattern already reappeared with the recent separation of energy transmission between the UK and the EU due to Brexit. The analysis and interpretation of multiplicity of solutions goes beyond this paper.

\section{Case Study: Results and Discussion}\label{s:Case_Study_Results}

\begin{figure*}[t]
  \begin{center}
  \subfloat{\includegraphics[width = 8.7cm]{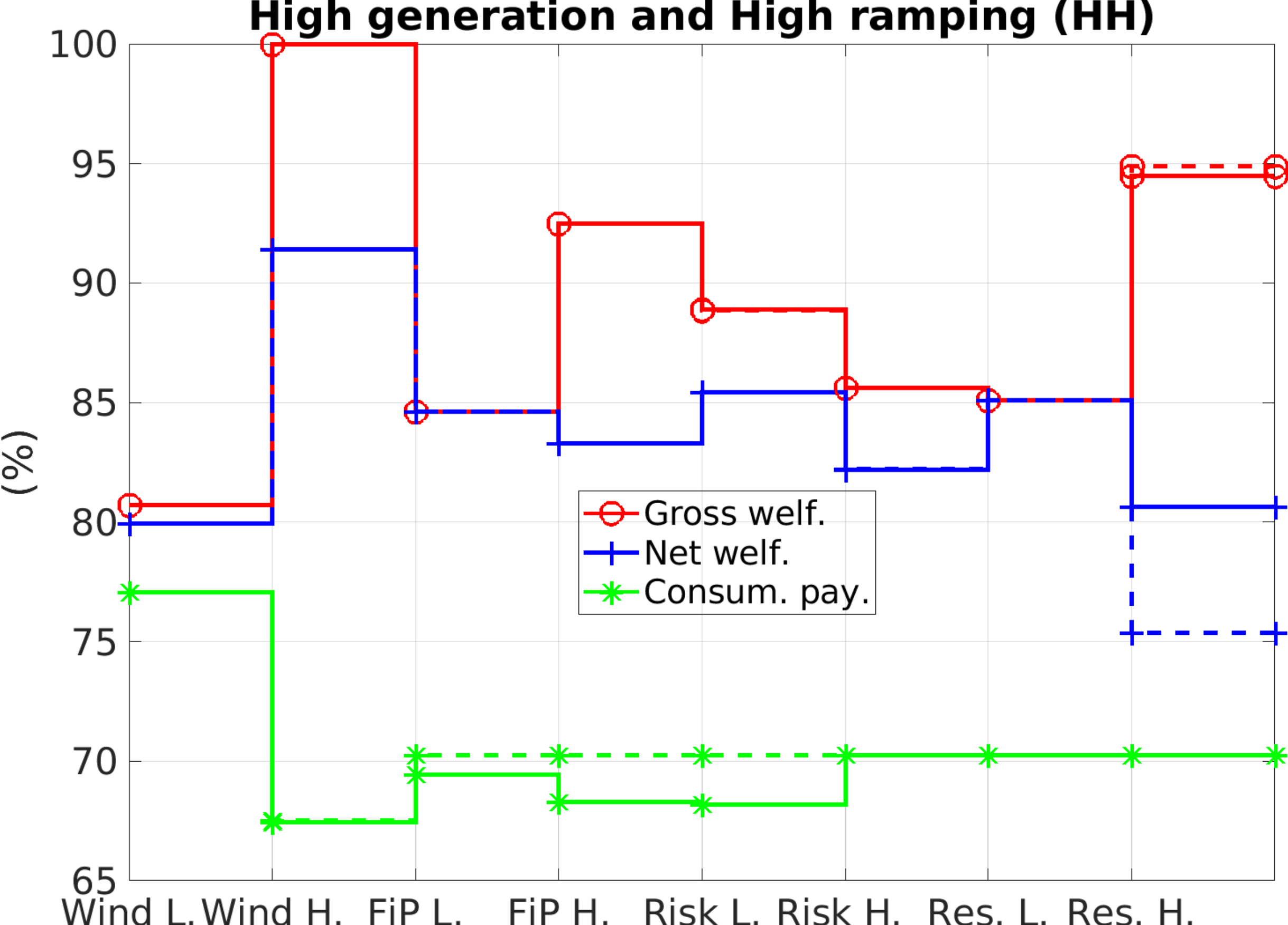}\label{fig:HH1}}
  \hspace{0.3cm}\subfloat{\includegraphics[width = 8.7cm]{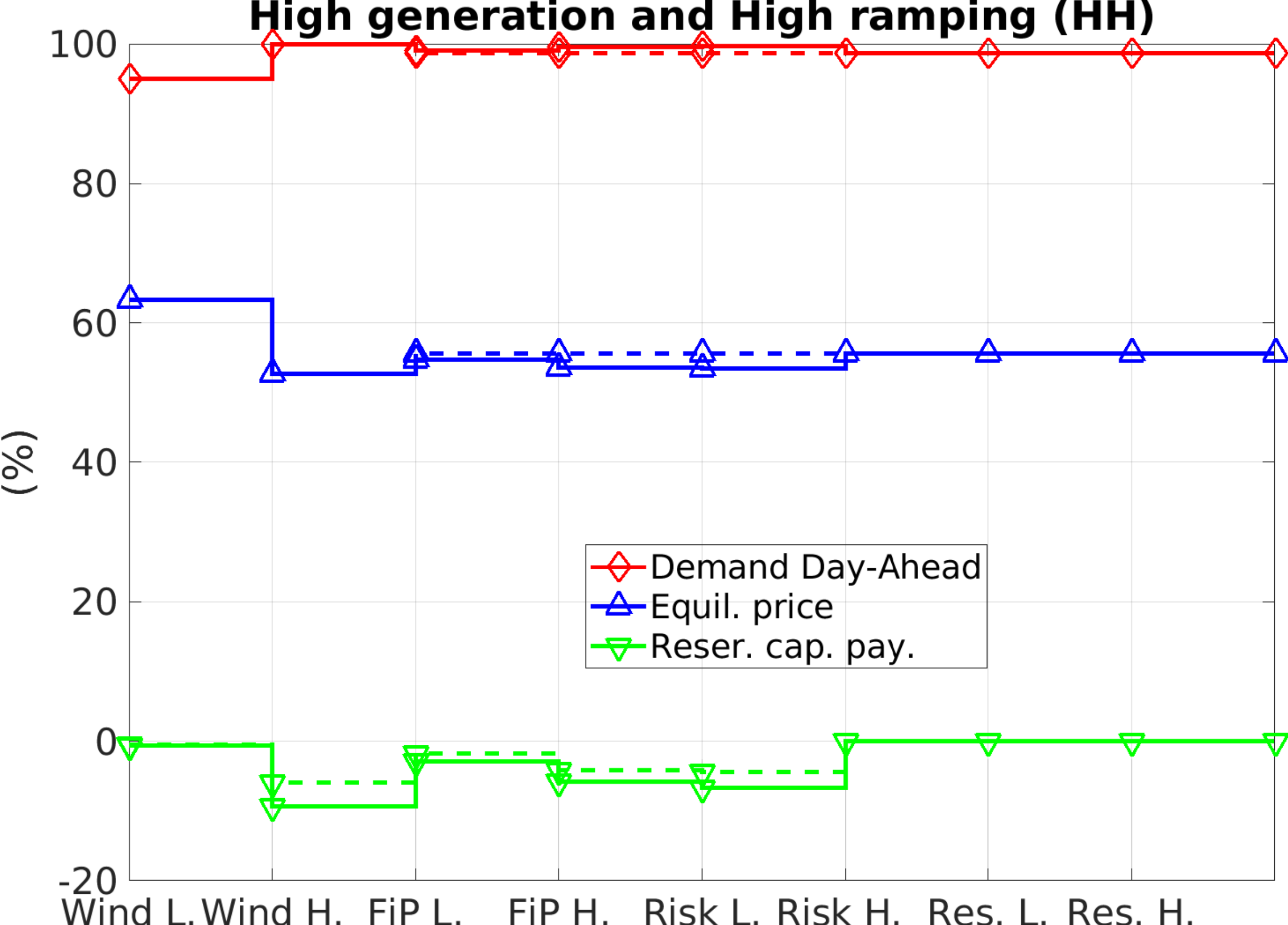}\label{fig:HH2}}
   \caption{Results for High generation and High ramping (HH). COM is in continuous line and EQM is in dashed line.}\label{fig:HH}
   \end{center}
\end{figure*}

Samples of results for the eight configurations listed on Table \ref{tab:ParametersValues} are depicted in 
Figs. \ref{fig:HH} to \ref{fig:VLH}. The configurations differ by the parameters ($\mu$, $FiP$, $\Psi$, $M_y$)  and are referred to by the column names in Table \ref{tab:ParametersValues}. To facilitate comparisons results are expressed in percentage of a base value, as explained below. COM and EQM results are respectively represented in continuous and dashed lines in each figure. The selected results and the base values are listed and briefly described below:

\begin{enumerate}
 \item (Demand Day-Ahead)\footnote{Names in parenthesis refer to the figures.} with base value 30120.39 MWh.
 \item (Equil. price), with base value 78.15 \officialeuro/MWh, that corresponds to the highest equilibrium price for the Low Generation configurations. Using that base value, the maximum value of variable cost for generators, $\max_g \{C_g\}$, is around 55.5\%. This means a (Equil. price)  lower than 55.5\% is an incentive for plant retiring.
 \item (Gross welf.), gross welfare, with base value 3552878.90 \officialeuro. It is computed as the objective function for COM both for COM and EQM.
 
 \item (Net welf.), net welfare, with the same base value as (Gross welf.), is obtained by subtracting the $FiP$ payment, $\sum_{w \in WT} y_w \!\cdot\! FiP$, from the (Gross welf.).
\item (Consum. pay.): consumer payment for energy, $(\Gamma_0 - \Phi_0 \!\cdot\! d)\!\cdot\! d$, with base value 1837531.10 \officialeuro. This payment goes to the generator.
\item (Reser. cap. pay.): payment for committed reserve capacities, $\sum_{g \in G} (ru_g \!\cdot\! (\overline{\kappa} - \overline{\gamma}) + rd_g \!\cdot\! (\underline{\kappa} - \underline{\gamma}))$, with base value 1837531.10 \officialeuro\hspace{0.05cm}, the same as for the (Consum. pay.), so both percentages can be added to get a global payment.

Because reserve is a service managed by the TSO in EQM, this payment goes from the consumer to the TSO, when it needs to incentivize the generator to provide reserve. Because the Spanish system includes an upper bound on reserve, we also introduce penalties, $\overline{\gamma}$ and $\underline{\gamma}$, on the generator in order to induce it to remain within these bounds. In that case, these penalties are levied by the TSO and rebated to the consumer as a reduction of fixed access charges. 
\end{enumerate}

A tentative motivation for the TSO's upper bound for reserve is briefly discussed here. Because of zero marginal cost, margins accruing from wind generation are higher than those from fossil fuel plants. This effect is further enhanced by the $FiP$. There is thus an incentive to bid wind instead of fossil plants in Day-Ahead (DA). The Real Time (RT) correction in case of discrepancies between DA forecast and RT realization mitigates this incentive, but to an extent that is difficult to foresee ex ante. There may thus remain an incentive to bid wind higher than expectation and to keep fossil capacities for reserve. The TSO may want to restrict this practice: it can do so by setting a reserve requirement and adding some interval (from 90 to 110\%) of that value for the generator to choose. This $[90,110]$\% is referred to as the TSO's interval in the following.

The common EU wisdom is that the electricity market is a commodity (energy) matter and that services (here reserve) are an other business. This is reflected in the separation of the Power eXchanges and TSOs. Market conditions where energy prices are insufficient for remunerating plants that are necessary for the functioning of the system, suggest exploring whether services (reserve) provided by the generators could constitute an other valuable source of revenue. This underpins the case study with the subsidiary question of whether co-optimization of energy and reserve could modify the value of the reserves. Three different situations can be identified:
\textit{i)} Reserve is available as an abundant byproduct of generation and hence has ``no value''; \textit{ii)} Reserve is excessive, it hits the upper value of the TSO's interval, and has a negative value. Increasing the upper bound to a level where it is no longer binding will extend the range where it has ``no value''; and \textit{iii)} The case of most interest is when reserve is effectively a scarce resource; it hits the lower value (90\%) of the TSO's interval and should be remunerated.

\subsection{High Generation Capacity (28.5 GW) and High Ramping Capability ($\Lambda = 0$): HH (Fig \ref{fig:HH})}\label{ss:Case_Study_HH}
 This case (HH) fully reflects the 2013 statement ``\textit{There's plenty of flexibility but so far it has no value}'' \cite{2013_Agora}. Electricity prices (Equil. price) are lower or equal to the highest fuel cost of the CCGT in most cases and hence will also not cover the fixed operating costs of all plants\footnote{Verifying that statement would require a full profile of wind availability (multi-period) and hence a more developed model,  our statement is thus only a judgmental comparison.}. The problem is structural and due to a legacy of flexible generation capacity that exceeds the need for reserve and generation. This is a short-term problem as the excess capacity will eventually retire. But it creates a stranded asset issue that needs to be managed: one needs an ordered exit in order to avoid a simultaneous dismantling of all the excess capacity. In contrast, and assuming that the unprofitable plants are retained through some special contractual arrangement, the consumer benefits from the disequilibrium: its bill (Consum. pay.)  is low and it receives a rebate on access charges (Reser. cap. pay.), this can obviously not last. The supply of reserve is excessive in almost all configurations with just a few cases where it remains within the TSO's interval. The following elaborates on this general situation: it  applies to both COM and EQM. 

Reserves hit the upper bound of the TSO's interval and are thus excessive in five configurations. They are signaled by negative Reserve Capacity payments (Reser. cap. pay.). 

The three other configurations have zero reserve capacity payment with their reserve remaining within the TSO's interval. They occur in case of risk aversion (Risk H.) and  high reserve requirement, $M_y=60$\%, which corresponds to an imprecise DA forecast of RT wind (Res. H. and Res. L.).  

The low electricity price and the excessive level of the reserve are two faces of a same coin, with the financial consequence that the latter reinforces the impact of the former. 
 
 COM and EQM only show small differences: demand in DA is higher and energy price (Equil. price) lower in a few COM configurations. Reserve is also more efficiently used in COM as shown by smaller payment from generators (Reser. cap. pay.) to TSO  reflecting less excessive reserve. (Gross welf.) are a bit higher in EQM, but (Net welf.) are identical. Existing differences are real but small. The striking feature of the situation is the dramatic disequilibrium in the generation system, that indeed prevailed in Europe before massive capacity impairment.

\begin{figure*}[t]
  \begin{center}
  \subfloat{\includegraphics[width = 8.7cm]{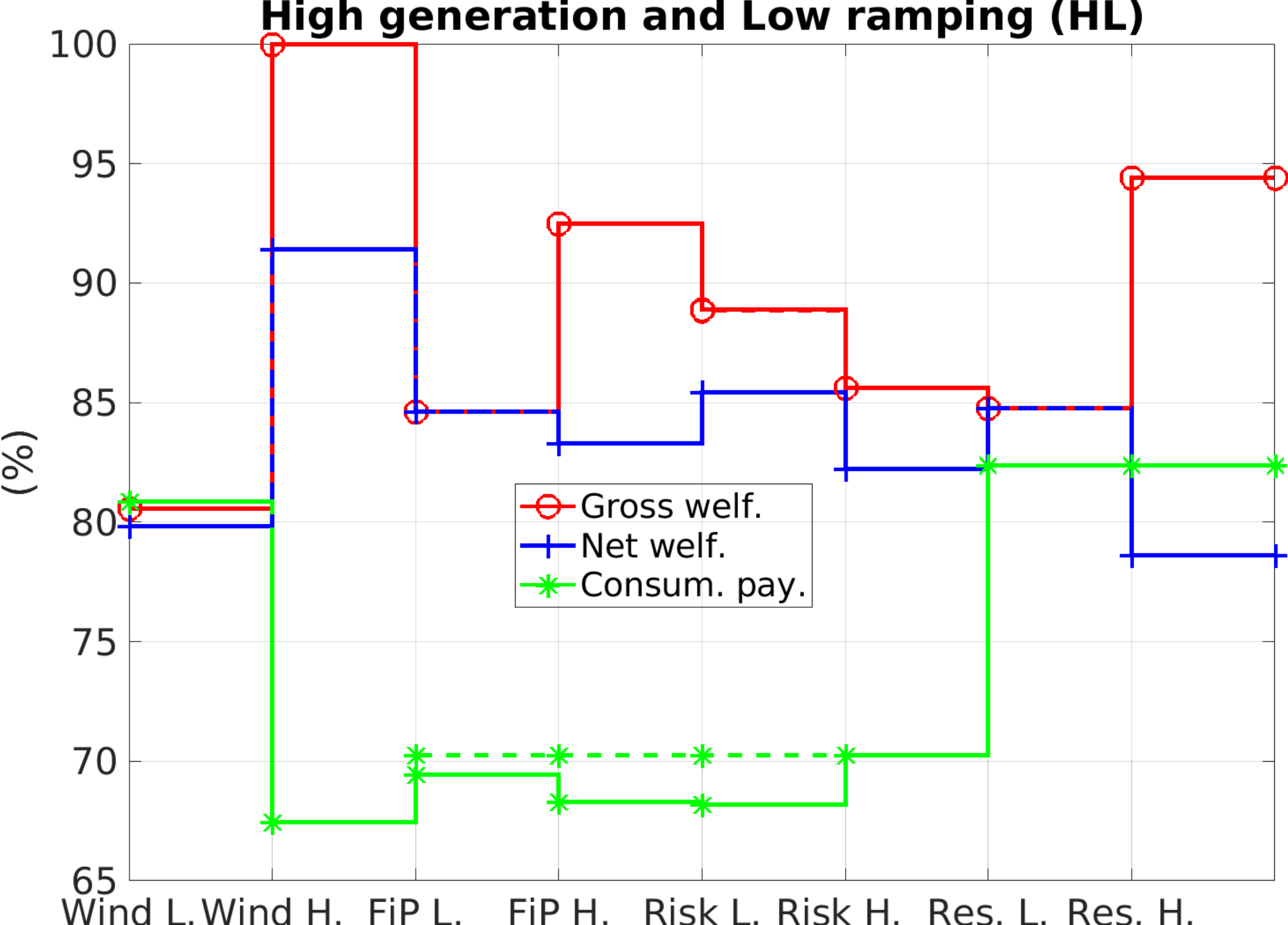}\label{fig:HL1}}
  \hspace{0.3cm}\subfloat{\includegraphics[width = 8.7cm]{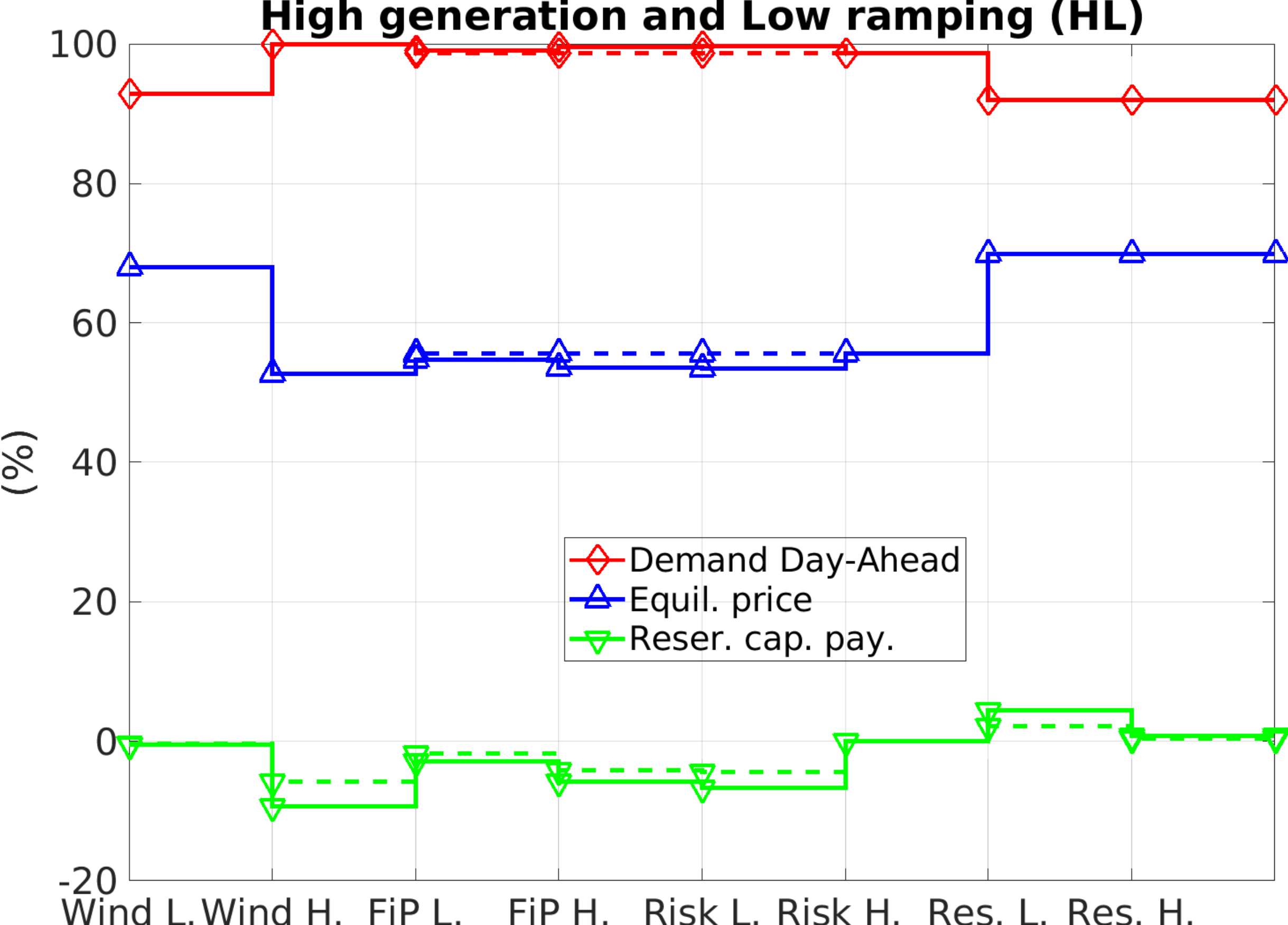}\label{fig:HL2}}
   \caption{Results for High generation and Low ramping (HL). COM is in continuous line and EQM is in dashed line.}\label{fig:HL}
   \end{center}
\end{figure*}

\subsection{High Generation Capacity (28.5 GW) and Low Ramping Capability ($\Lambda = 1$): HL (Fig. \ref{fig:HL})} \label{ss:Case_Study_HL}
This case (LH) slightly differs from the previous one (HH). It changes the value of energy and reserves and moderately increases the differences between the results of EQM and COM that remain limited to the same items: electricity demand, equilibrium price and reserve capacity payment. 
Electricity price (Equil. price) increases compared to HH  but remains generally too low to cover both fuel costs and fixed operating costs, except when demand for reserve increases because of imprecise forecast (high $M_y$ cases in Res. H. and Res. L.) or low wind (Wind. L.). 
Low wind  reduces demand because it implies higher fossil generation, with the aggravation that low ramping also contributes to the reduction of demand in DA (92.92\% in HL compared to 95.08\% in HH). This is neither a scarcity of capacity nor a market power effect: it is more profitable to use capacity for reserve of wind generation than directly for fossil generation. Recall that hindering that pattern has been mentioned above as one possible justification of the upper bound of the TSO's interval ([90,110]\% reserve requirement).

The reserve behavior is mostly analogous to the one in HH, Fig. \ref{fig:HH}, but a bit tighter. This is due to the lower flexibility that reduces the contribution of capacity to reserve. Generally speaking, reserve remains in the ``no value'' area with (Reser. cap. pay.) negative or zero. An exception occurs with high requirement for reserve capacity (Res.L and Res.H) and risk aversion high  (Risk H.) where (Reser. cap. pay.) is zero or positive. Reserve is now higher or equal to the lower bound of the TSO's interval. The energy price (Equil. price) in high reserve demand (Res. H. and Res. L.) comfortably exceeds the fuel cost of the plants, which should make them able to cover their fixed operating costs. The same is true in low wind (Wind. L.). The more efficient use of reserve in COM than in EQM observed in HH is mostly confirmed with reserve being now positive with high reserved demand (Res. H.), except for one case that deserves a particular attention. Configuration (Res. L.) with high demand for reserve shows a reserve capacity payment higher in COM than in EQM, which is an anomaly with respect to all other results.

We conducted a deeper investigation of (Res. L.), using a homotopy reasoning from configuration (FiP. L.) to (Res. L.). This amounts to parameterize the problem from the low to the high demand for reserve, $M_y = 15$\% $\rightarrow$ $M_y = 60$\%, (all other parameters remaining equal). The results show a discontinuous behavior for a demand of reserve $M_y$ around 46.5\% where PATH switches from an isolated equilibrium to an other one where it remains till $M_y = 60$\%. Note that the results remain within basic principles, in other words EQM does not dominate COM in terms of welfare, which would have contradicted basic economics. It is simply an illustration that the lack of co-optimization, that is the separate clearing of energy and reserve, creates uncertainty (and counter intuitive behaviors) in the outcome of the market and that these are worth further exploration. 

The general conclusion of these two cases (HH and HL), high generation with high/low ramping, is that except in the configurations of high reserve demand, and low wind one should expect some generation capacity to be retired in order to restore financial sustainability. The next cases examine this situation.

\begin{figure*}[t]
  \begin{center}
  \subfloat{\includegraphics[width = 8.7cm]{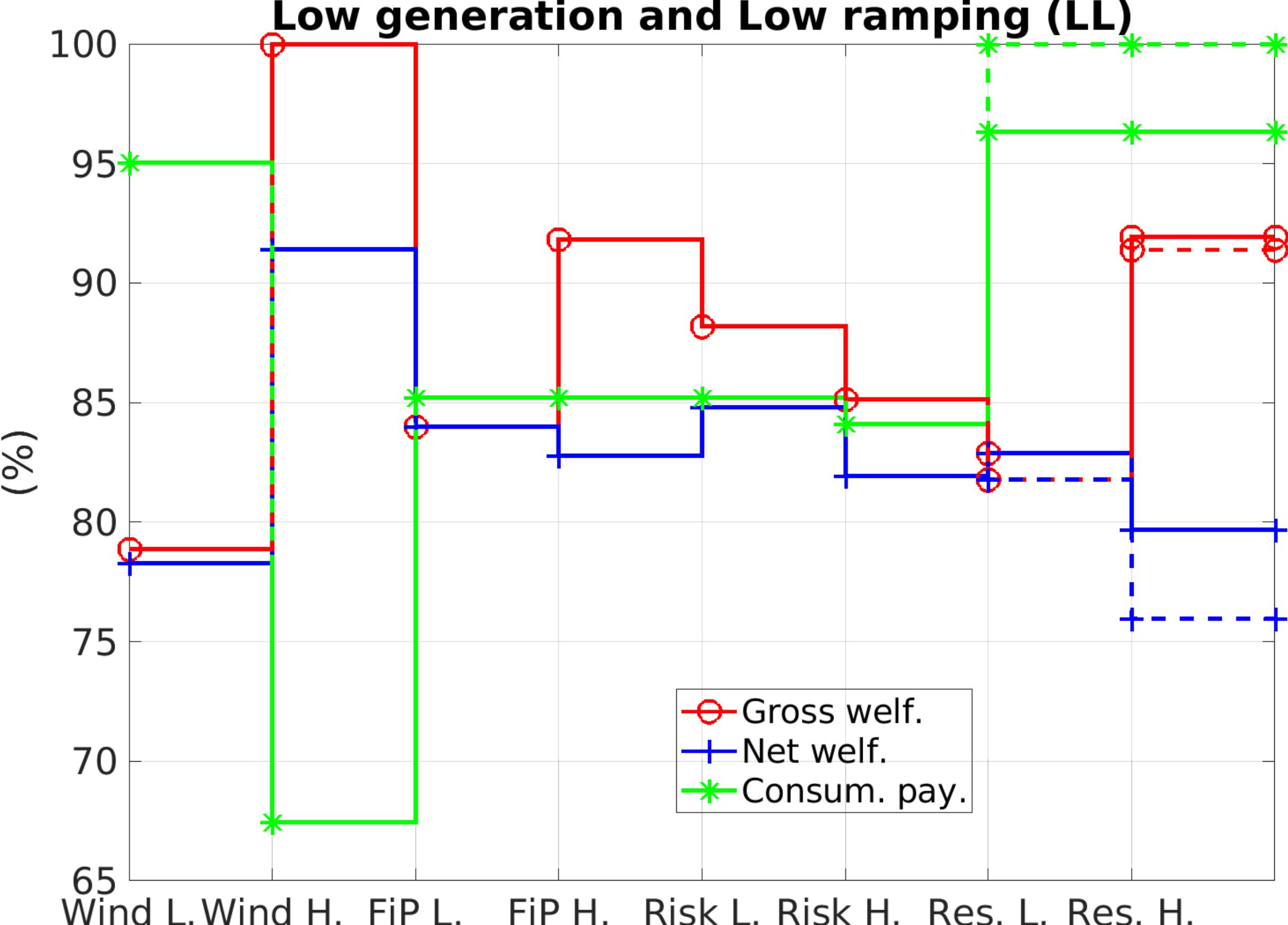}\label{fig:LL1}}
  \hspace{0.3cm}\subfloat{\includegraphics[width = 8.7cm]{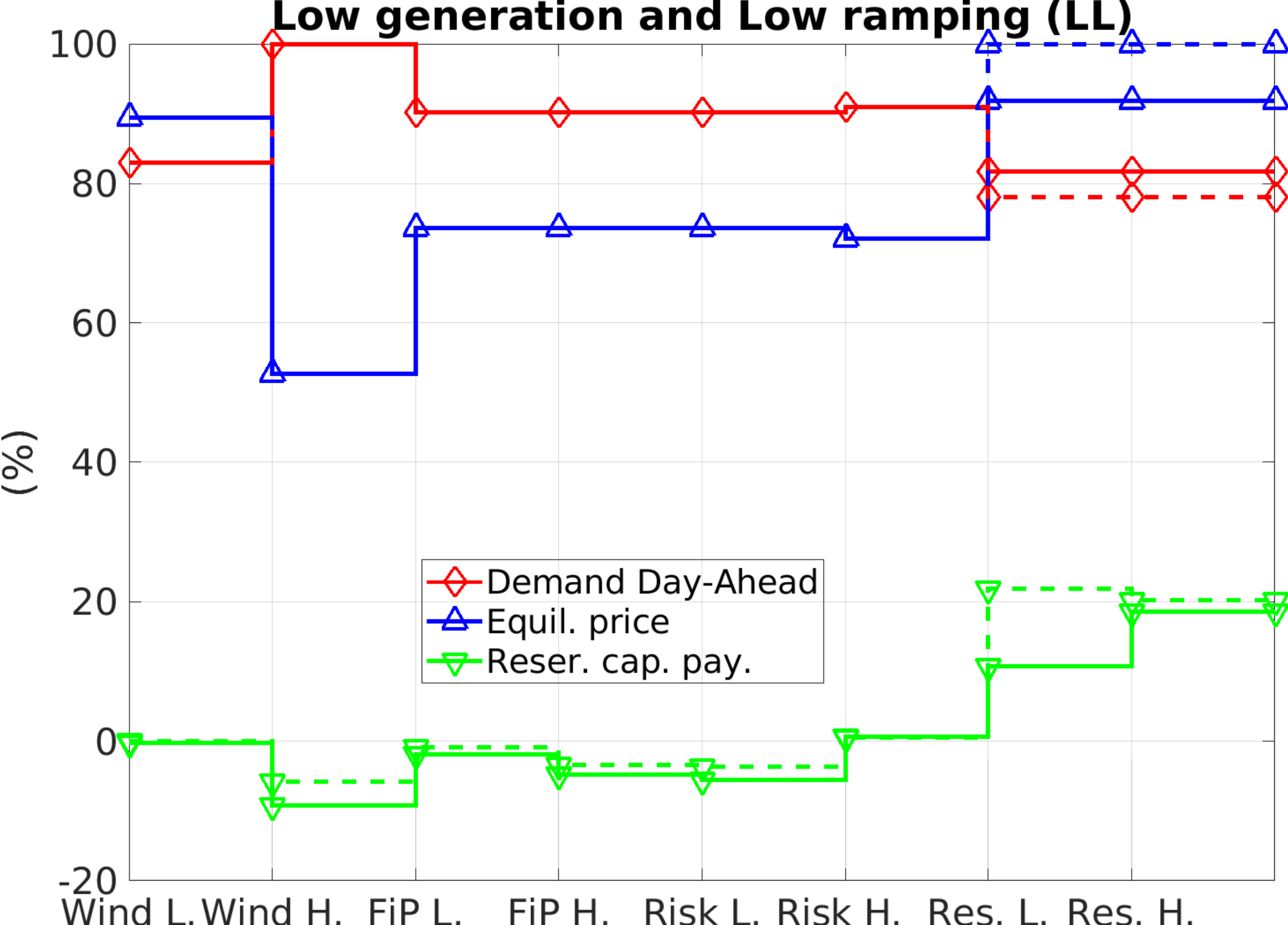}\label{fig:LL2}}
   \caption{Results for Low generation and Low ramping (LL). COM is in continuous line and EQM is in dashed line.}\label{fig:LL}
   \end{center}
\end{figure*}

\subsection{Low Generation Capacity (25.35 GW) and Low Ramping Capability ($\Lambda = 1$): LL (Fig. \ref{fig:LL})}\label{ss:Case_Study_LL}

A reduction of capacity is the expected logical consequence of the loss of profitability of generation observed in Figs. \ref{fig:HH} (HH) and \ref{fig:HL} (HL). This reduction should in principle increase the price of energy, as well as the revenue accruing from, now possibly scarce, reserves. This is effectively what happens in (LL), at least for energy: the price (Equil. price) is now sufficiently higher than the 55.5\% benchmark to suggest that it can support fuel and fixed operating costs. Inducing investment is obviously another matter. The only exception to this general finding occurs with high wind (Wind H.) where energy price is still a low 52.65\%. 
 Reserve also look better even though generally remaining abundant as observed by the profile of reserve capacity payment (Res cap. pay), which remain  negative or zero in six configurations, some of them associated to excessive reserves. Positive values occur for high reserve demand (Res. H. and Res. L.): the reserve capacity in those cases is effectively a payment from the consumer to the TSO,  and eventually to the generator that contributes to the value of the plant. 
 
 Other phenomena are worth mentioning. The payment for excessive reserve that reduces the value of capacities decreases compared to the case with higher generation capacity (HL): excess reserve diminishes when excess generation capacity decreases. Similarly the positive value of the reserve capacity increases the contribution to the plants profit, with (Reser. cap. pay.) becoming a non negligible fraction of the revenue of the plants when the demand for reserve is high (Res. L. and Res. H.). Also EQM and COM now perform differently, with COM significantly reducing the cost of reserve capacity. This has consequences on the electricity price and the payment for services by the consumers (both lower in COM than in EQM), implying a (modestly) higher welfare in COM. Co-optimization of energy and reserve thus appears useful if the demand for reserve is high. This suggests further reducing the capacity to test a possibly general scarcity of reserves, what is discussed next.

\begin{figure*}[t]
  \begin{center}
  \subfloat{\includegraphics[width = 8.7cm]{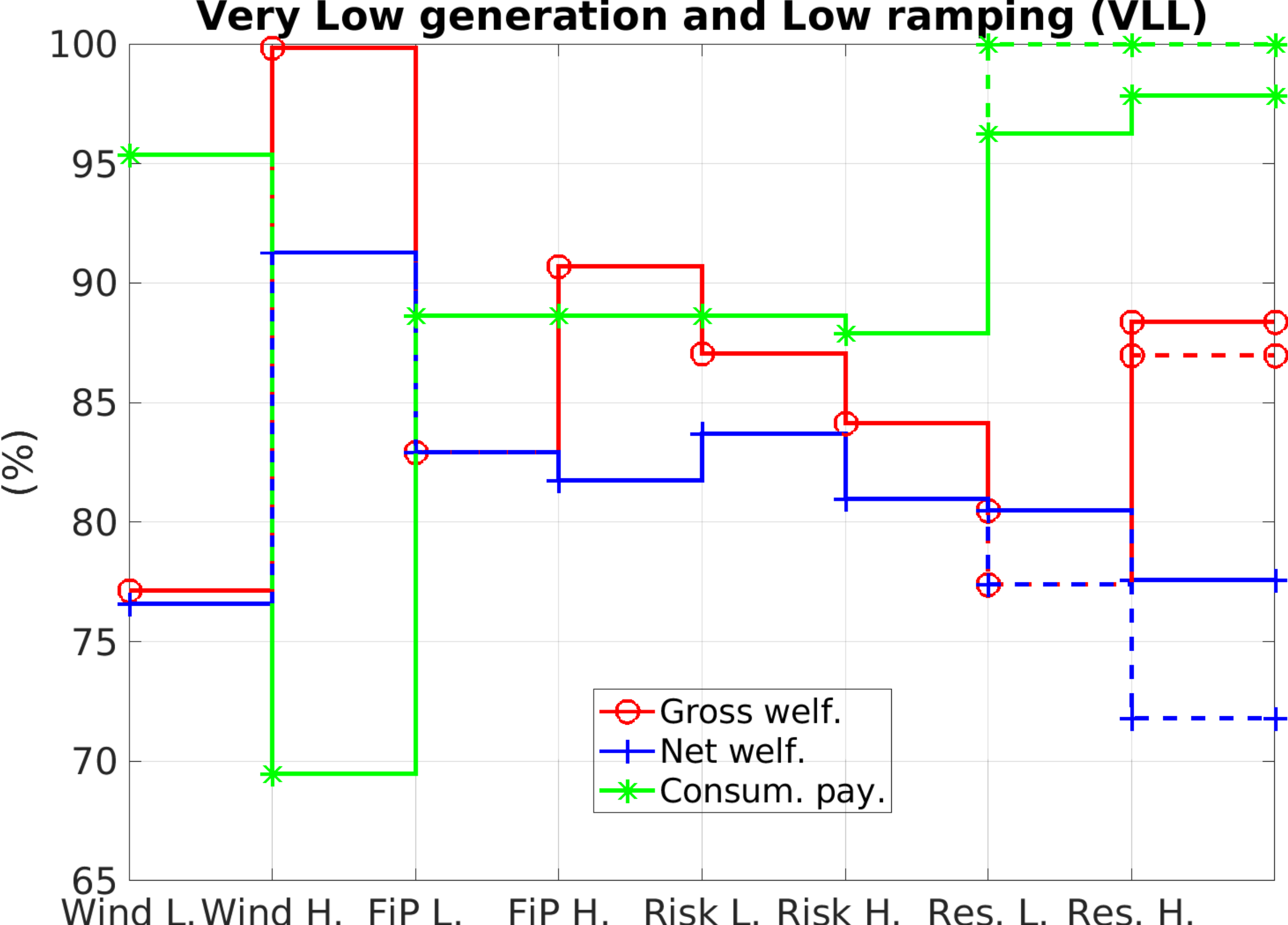}\label{fig:VLL1}}
  \hspace{0.2cm}\subfloat{\includegraphics[width = 8.7cm]{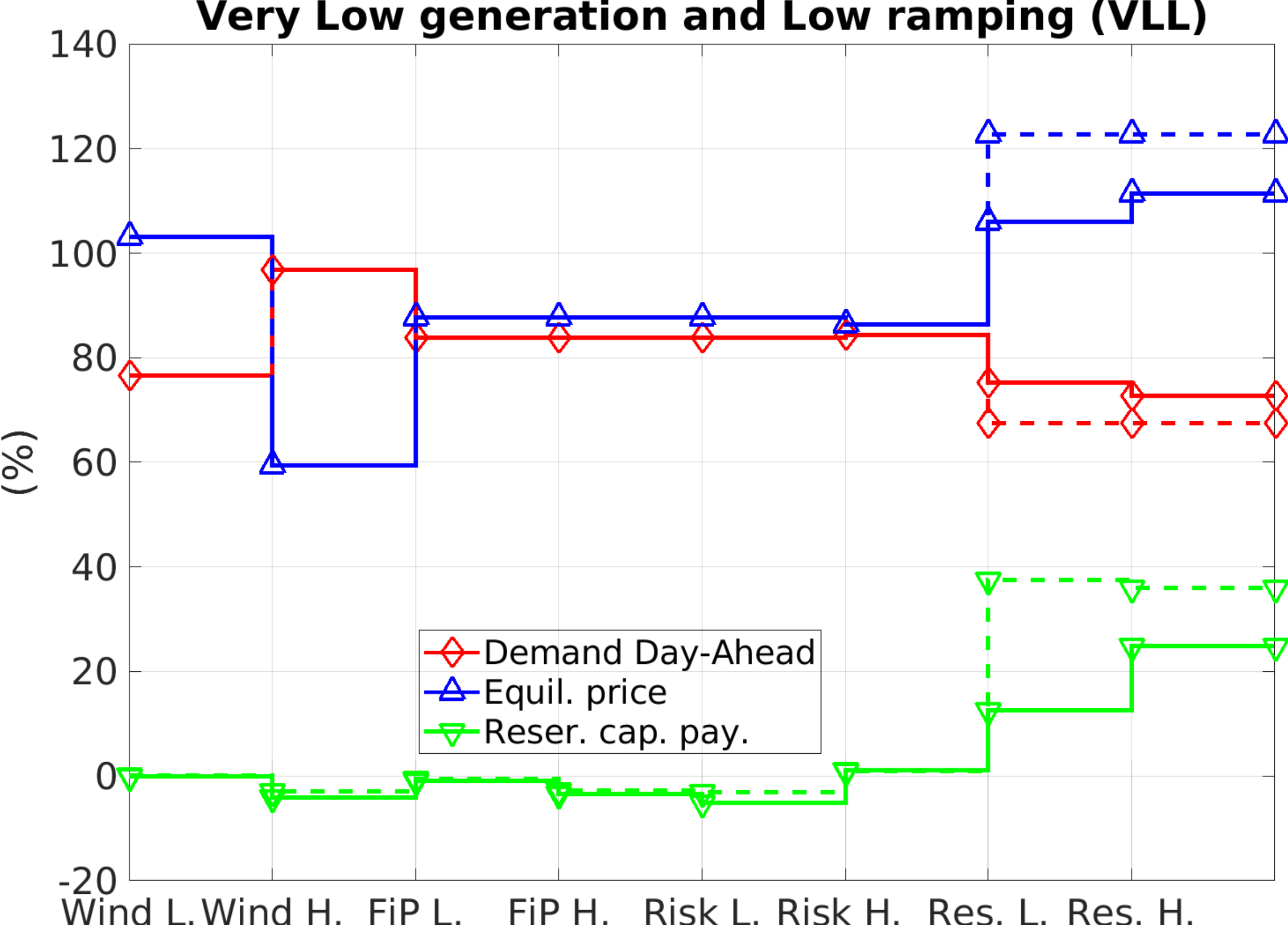}\label{fig:VLL2}}
   \caption{Results for Very Low generation and Low ramping (VLL). COM is in continuous line and EQM is in dashed line.}\label{fig:VLL}
   \end{center}
\end{figure*}

\begin{figure*}[t]
  \begin{center}
  \includegraphics[width = 8.7cm]{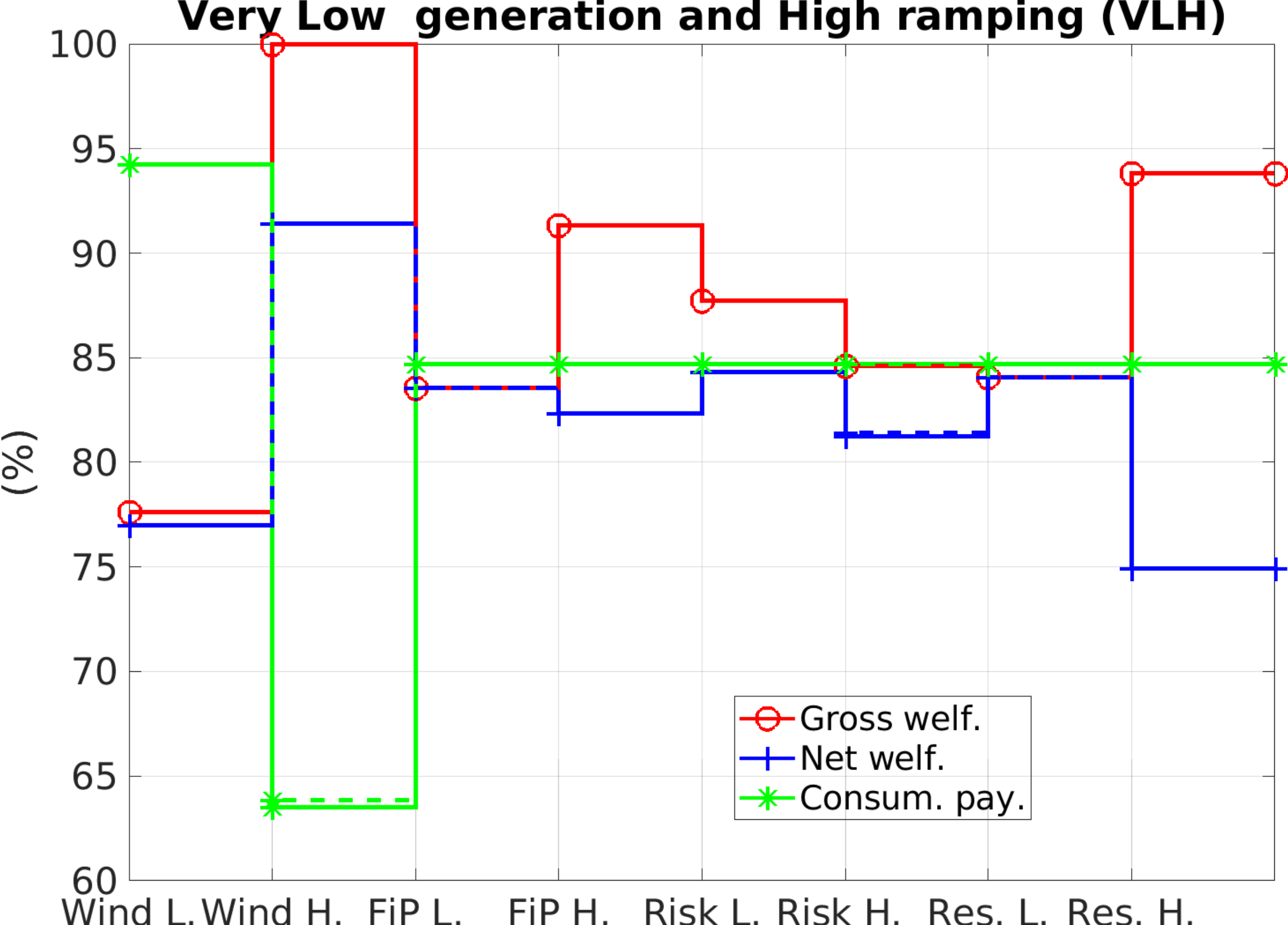}\label{fig:VLH1}
  \hspace{0.2cm}\includegraphics[width = 8.7cm]{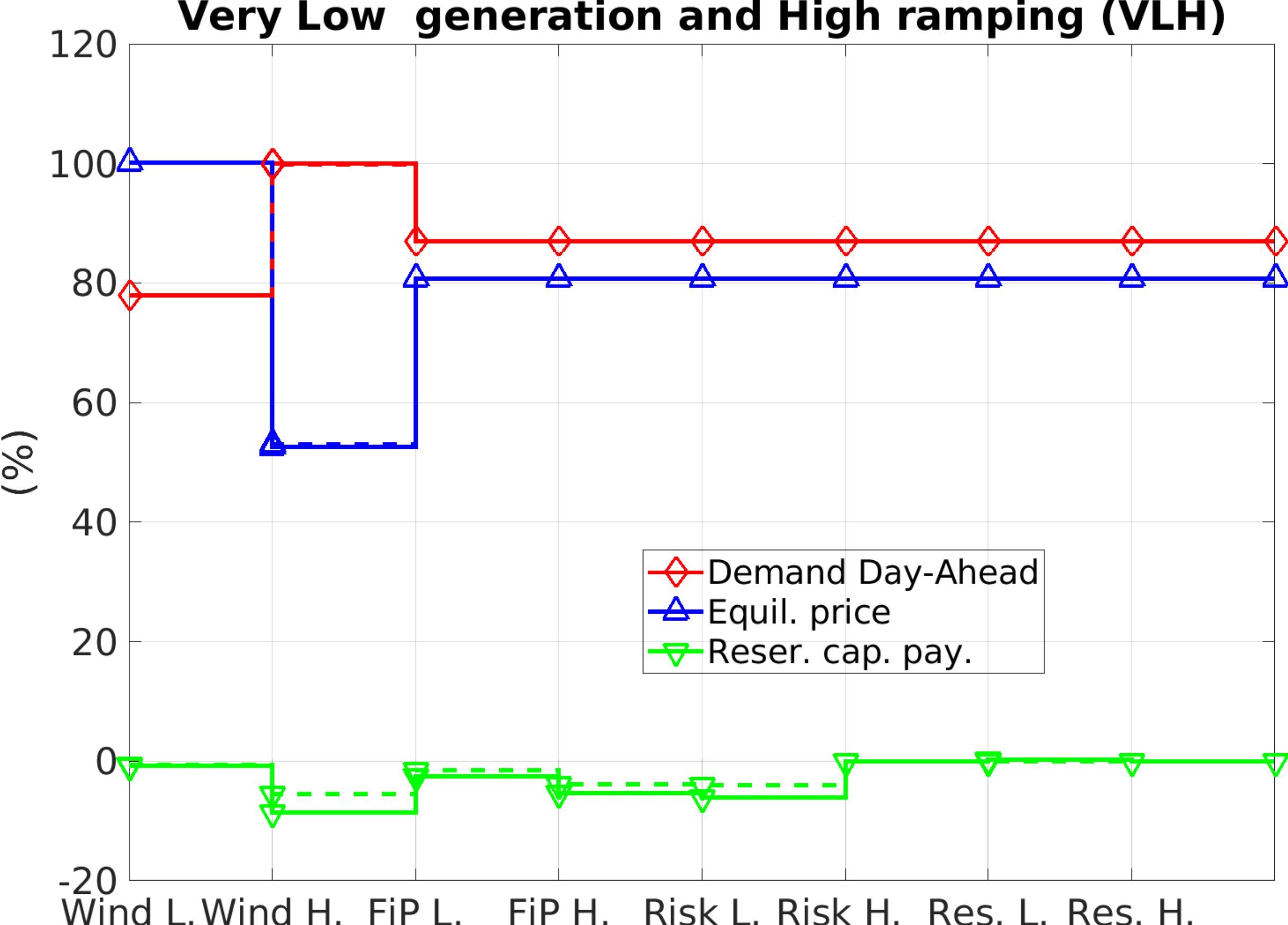}\label{fig:VLH2}
   \caption{Results for Very Low generation and High ramping (VLH). COM is in continuous line and EQM is in dashed line.}\label{fig:VLH}
   \end{center}
\end{figure*}

\subsection{Very Low Generation Capacity (23.35 GW) and Low ($\Lambda = 1$): VLL (Fig. \ref{fig:VLL})  /High ($\Lambda = 0$) Ramp. Capa.: VLH (Fig. \ref{fig:VLH})}\label{ss:Case_Study_VLL}

We conclude with two cases of a further reduction of capacity, both with very low dispatchable generation (23.35 GW), one with low ramping (VLL) Fig. \ref{fig:VLL}, the other with high ramping (VLH) Fig. \ref{fig:VLH}. Most of the phenomena observed in Fig. \ref{fig:VLL} for VLL are similar to those discussed for LL, Fig. \ref{fig:LL}, just a bit stronger. The price of energy (Equil. price)  increases and the negative value of reserve capacity payment (Reser. cap. pay.) persists, but decreases in the configurations (Wind L.) and (Risk L.). Reserve thus remains non constraining in these cases. The co-optimization of energy and reserve in COM improves the efficiency of the market by avoiding diverting too much capacity from energy to reserve. Also the share of reserve drastically increases in importance in the remuneration of plants with significant differences of efficiency between EQM and COM. 

A striking result is that these effects disappear with a high ramping capacity as shown in Fig.  \ref{fig:VLH} for VLH. The valuation of reserves and the role of market design are now completely different. The price of electricity (Equil. price) remains high enough to support fuel and fixed operating cost of the plants: this is a result of the lower capacity. But reserves loose all their values and return to the pattern depicted in Fig. \ref{fig:HH} for HH. The (Reser. cap. pay.) becomes highly negative in the configurations for different wind (Wind L., Wind H.), $FiP$ (FiP L., FiP H.), and risk neutral generators (Risk L.), and close to zero (or zero) with high risk aversion (Risk H.) and high reserve requirements (Res. L., Res. H.). This suggests a very unstable market for reserve, at least in this model. 

\section{Conclusion}\label{s:Conclusion}

The dramatic impact of wind penetration on power markets is now well recognized. Wind decreases the price of the commodity and damages the economics of conventional and flexible generators. While this may seem like a good step towards the energy transition, this creates stranded costs for plant owners. It also raises an issue of vulnerability of the energy market if flexible capacity is suddenly retired from the system, because it can no longer cover its fixed operating costs, and alternatives are not yet available. The phenomenon is now well understood; it was most flagrant in the period 2008-2013 in the EU, and recent documents show that it is still rampant. Possibly less understood is the extent to which the demand for reserve by the TSO for accommodating wind, can create a countervailing demand for services that could mitigate or substitute the pressure coming from low energy prices whether in the residual conventional system or its replacement. Still less understood might be the possible impact of market design on that countervailing demand. More specifically in this paper, the EU develops a strong renewable policy in a regime of separation of energy and services. Because there exists clear arbitrage possibilities between energy and reserve a relevant objective is to try to characterize the demand for reserve services in the current evolution of reduced fossil generation and whether the EU separation of energy and reserve has an impact on that demand compared to a co-optimization design.

This work attempts to provide some insight on this issue. Leaving aside transmission we construct two models that embed energy and reserve in two market organizations: EQM keeps energy and reserve separate and COM co-optimizes them. The models are not meant to represent a particular market, but try of satisfy some degree of realism by taking inspiration from a European situation. The analysis is conducted by referring to the (costly) evolution of fossil generation since the beginning of this century and in particular in the period 2008-2013.   


We find that there is no pricing power in reserve when generation capacity is important and fossil plants are unable to cover their fixed and variable operating costs on the energy market. Market design (COM or EQM) is also irrelevant. But a high demand for reserve due to wind combined with a reduction of economically redundant generation capacity restores plant profitability, reveals the value of reserve and the relevance of co-optimization at least in a basic situation of little flexibility capability. The revenue accruing from reserve significantly contribute to the viability of flexible plants in COM and EQM, but because of a more efficiently use of reserves COM decreases the price to the consumer and increases demand and welfare. But these advantages look fragile to an increase of reserve capability. Increasing the reserve potential of existing plant could quickly decrease their value both in COM and EQM and be possibly self defeating for an investor.

This result obviously requires further analysis. This should be combined with the relaxation of important simplifications made in the representation of the EU system idiosyncrasies that could only be removed at the cost of considerable technical difficulties. This will be treated in a future work.

\ifCLASSOPTIONcaptionsoff
 \newpage
\fi

\bibliography{References_VT_v10}  
\bibliographystyle{IEEEtran}


\begin{IEEEbiographynophoto}{Yves Smeers} 
 received the M.S. and Ph.D. degrees from Carnegie Mellon University, Pittsburgh, PA, in
1971 and 1972, respectively.\\
He is currently a Professor Emeritus and a member of the Center for Operations Research and Econome\-trics, at Universit\'e Catholique de Louvain, Louvain-la-Neuve, Belgium.
\end{IEEEbiographynophoto}
\begin{IEEEbiographynophoto}{Sebasti\'an Mart\'\i{}n} 
(S'08, M'14) was born in M\'alaga, Spain, in 1980. He graduated from University of Granada, Spain, as Civil Engineer in 2003. He received an Industrial Engineer and a PhD degrees from University of M\'alaga, Spain, in 2007 and 2014 respectively.\\
He is currently a teaching assistant at the University of M\'alaga. His research interests include: optimization theory, methodologies for teaching, and economics of electric energy systems.
\end{IEEEbiographynophoto}
\vspace{-1cm}\begin{IEEEbiographynophoto}{Jos\'e Antonio Aguado} 
(M’01) received the Ingeniero El\'ectrico and Ph.D. degrees from the University of M\'alaga, M\'alaga, Spain, in 1997 and 2001, respectively.\\
Currently, he is full Professor and Head of the Department of Electrical Engineering at the University of M\'alaga. His research interests include ope\-ration and planning of smart grids and numerical optimization techniques.
\end{IEEEbiographynophoto}
\end{document}